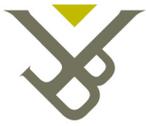
Vrije Universiteit Brussel

**N. Rons, A. De Bruyn and J. Cornelis**

N. Rons is Coordinator of Research Evaluations and Policy Studies, Research Coordination Division, Research & Development Department, Vrije Universiteit Brussel (VUB), B-1050 Brussels, Belgium; E-mail: nrons@vub.ac.be.

A. De Bruyn is Collaborator of the Research Co-ordination Division, Research & Development Department, Vrije Universiteit Brussel (VUB), B-1050 Brussels, Belgium; E-mail: ardebruy@vub.ac.be.

J. Cornelis is Vice-Rector for Research, Vrije Universiteit Brussel (VUB), B-1050 Brussels, Belgium; E-mail: vice-rector.onderzoek@vub.ac.be

**Research Evaluation per Discipline: a Peer Review Method and its Outcomes**

***This paper describes the method for ex-post peer review evaluation per research discipline used at the Vrije Universiteit Brussel (VUB) and summarizes the outcomes obtained from it. The method produces pertinent advice and triggers responses - at the level of the individual researcher, the research team and the university's research management - for the benefit of research quality, competitivity and visibility. Imposed reflection and contacts during and after the evaluation procedure modify the individual researcher's attitude, improve the research teams' strategies and allow for the extraction of general recommendations that are used as discipline-dependent guidelines in the university's research management. The deep insights gained in the different research disciplines and the substantial data sets on their research, support the university management in its policy decisions and in building policy instruments. Moreover, the results are used as a basis for comparison with other assessments, leading to a better understanding of the possibilities and limitations of different evaluation processes. The peer review method can be applied systematically in a pluri-annual cycle of research discipline evaluations to build up a complete overview, or it can be activated on an ad hoc basis for a particular discipline, based on demands from research teams or on strategic or policy arguments.***

**Context & Goals**

Research evaluations are conducted with different goals and at different governance levels. Goals set by the university's research management (e.g. internal quality assessment, definition of spearhead research groups, evaluation of the PhD support and training processes, ex-ante and ex-post project evaluations) may be very different from those set by national authorities (e.g. foundation or justification for university funding, mapping of the global research potential, rationalizing the higher education sector).

In its regulations[1] the Flemish government imposed a periodic evaluation of all research at all universities of the Flemish community in Belgium. The universities can independently choose or develop their own research evaluation tools, so that these can optimally serve specific goals in research quality. In response to the government requirements, the Research Council of the Vrije Universiteit Brussel (VUB) installed in 1996 a systematic, two-scale, ex-post evaluation procedure for the research programs and processes within each *scientific discipline*[2] and for each *research team* whose research is primarily compliant to the discipline description. The scale of an evaluation influences the choice and the feasibility of evaluation methods, and thus the nature of the results that can be obtained. The method described here fulfills the government requirements towards the university and serves the university's research policy at its typical management levels. On one hand the evaluations are carried out at the scale of a *scientific discipline*, a granularity well suited for management at the departmental and university levels. On the other hand the subjects of evaluation are the smallest groups in the university's structure, namely *research teams*, which allows for local remediation at team level and up to the level of the individual researcher.

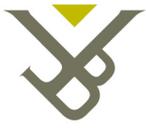

At the time these ex-post evaluations were introduced at the VUB, previous experience in research evaluation within the academic world primarily concerned ex-ante evaluations of project or program proposals (*CRE*, 2000). The ex-post evaluations procedure described in this paper complemented the existing VUB-system of ex-ante assessments of *personal research records* and *individual project evaluations* (Cornelis, 2005) with a system for the evaluation of *research programs and processes in all scientific disciplines*. The main goal of the ex-post evaluations is to obtain clear, pertinent and useful advice, aimed at remediation and research quality improvement wherever possible. In view of this goal, an original method was designed for the evaluation of research teams, grouped per discipline. The general results for each discipline are communicated to the Research Council and to the Board of the University, in a public report as required by the Flemish government. The more detailed reporting per team remains confidential to ensure that the experts do not feel refrained from formulating their advice and critical observations. For the same reason no *automatic* consequences are linked to the results. The commitment of the university management is that these evaluations are thorough but "friendly", leading to internal and concerted initiatives for remediation. A lot of effort has been spent on making this last commitment credible within the research community.

**Peer review**

All types of evaluation methods have their own advantages, limitations and typical bias. In view of the goal that was set, the Research Council chose for a qualitative, *peer review* based method. The Research Council considers peer review as the most effective policy instrument for *quality improvement* and the best way to obtain a thorough and useful view on the research *processes and quality*. Furthermore, the peer review principle can be applied to *all disciplines*. This is not always the case for quantitative models, such as bibliometrics, which are always limited to the coverage of the database used (van Raan, 2005; Nederhof, 2006)[3]. Moreover, research *process* information, which is crucial for remediation purposes, cannot be obtained solely by analysis of output. Output parameters (publication/citation or other) often ignore or underestimate more subtle aspects of research quality and culture, such as the *training* aspects and *international embedding* of young researchers, the research culture in which *students* are immersed, the *impact* of research on society, economy, culture, government policy, development aid, the scientific literacy of citizens and the non profit sector. The choice for peer review is supported by international studies, showing it is the most cited evaluation method for research quality (Jongbloed et al., 2000; *CRE*, 2000). The academic community has since long been using peer review (Rigby, 2003). Today it is used by well-established funding agencies such as the National Institutes of Health (NIH) where it evolves over time to address new realities (Scarpa, 2006). The more recently founded European Research Council (ERC) will use peer review panels to assess applications in all fields of science, as advised by an expert group (Mayor et al., 2003). More specifically in the Flemish context, the choice for peer review is in agreement with the conclusions of an external committee which evaluated the universities' research policy for the Flemish Interuniversity Council (VLIR) in 1996 (van Duinen, 1996).

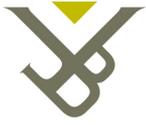

While peer review is generally accepted as the principal method to evaluate research quality, its sensitivity to bias is well known. Critical observations on peer review, performance measures, impacts of evaluations and experiences from many other evaluation exercises were taken into account for the design of the method in 1996-1997 (among many other contributions: Martin 1996; Luukkonen, 1995; OECD, 1997; Kostoff, 1997a; 1997b; Cozzens, 1997). Where possible, mechanisms to avoid known undesired effects were included in the method's design, increasing its reliability (see next section and Table 1). Peer review related criticisms are often set in a context of decisions for funding or publishing. They include policy issues such as uncoupling decision power from accountability, bias and limited selective power (Chubin & Hackett, 1990), the influence of chance, scales and budget constraints (Cole et al., 1981; Langfeldt, 2001). An analysis of peer review in its different forms and functions as a social phenomenon, aiming to inform in view of the development of a review system, is given by Hackett & Chubin (2003).

After a sufficient number of evaluations had been carried out the reliability was further investigated together with results from another Belgian university, the Universiteit Antwerpen (Rons & Spruyt, 2006). Discussions on peer review reliability, methodology and acceptance and performance measures in general, continue to be monitored for new elements which could further improve the developed method (e.g. Luukkonen, 2002; Tijssen, 2003; Langfeldt, 2004; Brown, 2004; Moed, 2005).

**Principles, Pitfalls & Precautions**

The main goal, namely to obtain clear and pertinent advice, determined a series of principles to be met by the method.

Good advice can only be given by *experts* in the field who are able to compare the performance to an *international* reference level. Their advice must be complemented with a correct interpretation in the *specific context* of the discipline. The experts should not feel refrained in any way to give their *honest and unfiltered opinion*. They must receive the *data* needed to form their opinion. All teams must be *treated equally*. Teams with *specific profiles*, e.g. focusing on innovative or interdisciplinary research, must not be disadvantaged. The resulting assessment should be *coherent* over all material evaluated. Sufficient *time* is needed to collect and clearly present the necessary information and to study the material and formulate advice. Misinterpretations are to be avoided and remaining questions should be resolved in *dialogue* format. Recommendations should be *selectively reported* to the appropriate governance levels to allow them to play *their* role in remediation.

These principles constitute the framework around which the method was developed. They imply working with *panels* of experts (coherence, anonymity), who are *visiting* the university (dialogue). Table 1 presents an overview of precautions taken in order to meet the principles set, and thus also of potential problems or pitfalls inherently linked to peer review, in this specific context of panel evaluations. This set of precautions also contributes to a higher acceptance by the researchers, as they tackle many of their criticisms, concerning mistakes, bias, superficiality, overlooking evidence, not taking all factors into account, underestimating originality, assessment based on one opinion only (Luukkonen, 1995).

**Table 1. Principles, Pitfalls & Precautions**

| Principles | Pitfalls | Precautions |
|---|---|---|
| 1. Domain Specificity | - Unadapted data / criteria / advice | - A coordinator from the field ensures that relevant data are collected and that comments are correctly situated for the particular domain. |
| 2. Expertise | - Mismatched competences | - A coordinator from the field composes a panel of experts competent in the fields concerned and experienced regarding evaluation. |
| 3. Independence & Objectivity | - Various kinds of bias | - The teams can suggest and reject experts, avoiding cognitive and negative bias.<br>- Connections with suggested experts are verified, avoiding positive bias.<br>- In the panel & meeting formula peers can supervise each other, discouraging the expression of bias.<br>- Each expert reviews all teams, limiting the influence of personal bias.<br>- The panel is composed of international experts, avoiding partiality due to a same national context. |
| 4. International Level | - Unchallenging or unequal standards | - The international panel composition ensures that performances are assessed based on world standards. |
| 5. Honest and Unfiltered Opinions | - Refrained advice | - Confidential reports aimed at concerted remediation (no *automatic* consequences) encourage open advice from experts and open presentations from the teams. |
| 6. Qualitative Data & Assessment | - Lack of strategic / background / contextual information | - The evaluation files contain a description of research activities, a SWOT analysis and personnel and teaching data, fully enabling the experts to assess the research performance and process.<br>- The meeting formula allows teams and research administration to informally provide the experts with extra contextual or more recent information. |
| 7. Uniform & Fair Treatment | - Superficiality<br>- Not all factors considered | - Uniform evaluation files and evaluation forms ensure that the same aspects are considered for all teams.<br>- One expert leads each team's assessment, ensuring that each team's performance is looked at in detail.<br>- Reports per team are uniform and balanced, as ensured by the coordinator's supervision. |
| 8. Special Attention for Specific Research Profiles | - Underestimation of innovative / interdisciplinary research | - Explicit attention from the experts is asked for innovative and interdisciplinary research.<br>- The meeting formula confronts different standards and perspectives. |
| 9. Coherence & Reliability | - Assessments based on one opinion / on different standards | - Every expert reviews all teams in view of a uniform and balanced report.<br>- For each team, priority (a higher weight) is given to ratings from reviewers with higher expertise in its field. |
| 10. Thorough, yet Timely | - Incomplete data / advice<br>- Outdated results | - A flexible timing is kept in function of good quality in all phases.<br>- The global time frame is restricted to one year for timely results. |
| 11. Efficient, Well Informed Dialogue | - Mistakes / misinterpretations | - Overviews of scores and comments allow experts to identify aspects to be clarified.<br>- Discussions during the meeting between experts, teams and university management allow clarifying remaining questions and avoiding misunderstandings. |
| 12. Concise & Selective Reporting | - Insufficient clarity in view of actions at different policy levels | - A global, public report on the discipline informs the institutional research policy management.<br>- More detailed, confidential reports per team inform each team leader, the dean of the faculty and the rector and vice-rectors. |

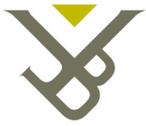

Each of the aforementioned principles is revisited here below and analyzed in more detail to illustrate how the goals associated to it are achieved. Cross-references illustrate how many aspects are linked and how choices aimed at one objective can automatically have an effect on other aspects. While the design of the method took place in 1996-1997, also some later contributions, from evaluation as well as policy experts, are referred to.
The next section gives more information on the practical organization of the evaluation.

1. **Domain Specificity**: The coordinator, appointed by the Research Council in concertation with the research team leaders, supervises all activities and ensures that the specificity of the scientific discipline is taken fully into account. If necessary, the coordinator can modify the standard procedure or documents according to the specific research culture of the discipline. The coordinator is a well-established researcher from the discipline but not involved in the evaluated research. A coordinator from a university in the same country has the advantage of knowing the national context and funding system.

2. **Expertise**: The choice of highly qualified, active peers is the most important factor for a successful research evaluation (Koetecky, 1999). The expert panel's expertise needs to cover the research fields of all teams. This is ensured by the coordinator (see 'Domain Specificity') who composes a panel of independent and experienced experts, for which the research teams can provide suggestions. The panel typically contains about as many experts as there are teams. As much as possible, the panel includes experts who have ample experience with evaluation procedures and who gained an overview of a larger part of the entire field.

3. **Independence & Objectivity**: The selection of experts is a crucial phase that may influence outcomes (Horrobin, 1990). The experts have to formulate an independent opinion, unbiased by e.g. collaborations and competition (see 'International Level'). Bias and fairness in peer review has been amply discussed in the literature, often in the context of the evaluation of manuscripts. Advice may be subject to professional or personal bias, related to interests or cognitive constraints (Langfeldt, 2004). As a precaution avoiding positive bias, links between teams and suggested experts are verified and discussed by the vice-rector for research and the coordinator. As a precautionary measure avoiding negative bias (rarely made use of), the teams can reject an expert if they can demonstrate why he/she cannot be expected to give objective advice. Furthermore, the panel and meeting formula enables the experts to supervise each other, discouraging the expression of personal predispositions. If bias does occur, its influence is limited to some extent by the fact that every expert reviews all teams (see 'Coherence & Reliability').

4. **International Level**: The teams' performances need to be compared to an international reference level. Therefore the panel contains primarily foreign experts and the evaluations are conducted in English. The international panel composition also avoids partiality that may result from involvement in a same national context (Koetecky, 1999; Stoicheff, 1999). An exception with respect to the international composition is made when the nature of the discipline (e.g. Law) requires national experts or another language. Under no circumstances may experts belong to the university. If a particular team's research output is mainly in another language, the panel includes an expert in its domain who has sufficient knowledge of this language.

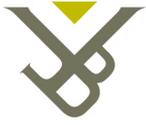

5. **Honest and Unfiltered Opinions**: The experts should not feel refrained and should state their opinions freely. When asked to publicly comment on the different research activities of a unit, in a report intended to affect the allocation of funding, the experts' advice may not be as clear as hoped for (Hämäläinen, 2000). Therefore, from the start, the aim of the evaluation project, the way the results will be used for concerted remediation (no *automatic* consequences), and the fact that results for individual teams remain confidential, are made clear to them. This information is also given from the start to the teams, so that they in turn are not refrained from presenting all relevant information.

6. **Qualitative Data & Assessment**: The experts have to receive all relevant information needed to form their opinion, including clearly described research results and directions (J. Koetecky, 1999). Data are presented in a qualitative way and interpreted by the experts based on their experience. The evaluation files contain an overview of research activities from the preceding five years, a period long enough to present a correct picture (limited risk of important elements falling out of the time frame) yet not too long so that the evaluations are based on recent information. All publications and projects are listed, ordered qualitatively in categories. Data on personnel and teaching load offer the necessary background for a correct interpretation of research performance. Descriptive paragraphs related to research management enable the experts to give advice on the quality of the research process. Elements the experts particularly appreciate are the core publications and the section concerning strengths, weaknesses, opportunities and threats (SWOT analysis). Complementary to the evaluation files, the meeting formula allows teams and research administration to informally provide the experts with extra contextual or more recent information (see 'Efficient, Well Informed Dialogue'). The experts do not only give scores but are explicitly invited to make maximal use of the possibility to provide comments.

7. **Uniform & Fair Treatment**: All teams need to be treated equally. Evaluation files of uniform layout and content ensure that the same information is available for each team (Table 2). Evaluation forms ensure that for each team the same aspects are considered (Table 3). In order to limit the efforts, each expert looks at the team in his own field with special attention. This attribution of the responsibility for a particular dossier to each expert ensures that each dossier gets the same attention and avoids that experts count on work done by other panel members, which is one of the group effects described by Langfeldt (2004). Yet, every expert looks at all documents so that a fair and consistent report can be made (see 'Coherence & Reliability') in which the anonymity of the experts can be guaranteed, in the sense that all advice is formulated by the panel as a whole. Finally, supervision by the coordinator ensures uniform and balanced reports per team.

**Table 2. Standard content of the evaluation files**

| Descriptive sections: | Data & overviews: |
|---|---|
| Preamble (a short presentation of the discipline) | B. Overview of the scientific activities |
| A. Presentation of the team | I. Publications |
| I. Introduction | II. Research projects |
| II. Research topics | III. Other scientific accomplishments (awards, promoters of PhD theses, memberships, …) |
| III. Most important research results | C. Financial means and personnel |
| IV. Past, present and future activities (objectives and strategy) | I. Personnel — Overview |
| V. Strengths and weaknesses / threats and opportunities | II. Personnel — Details |
| | III. Teaching load |
| VI. Scientific and social relevance of the research | IV. Most important sources of funding |
| VII. Short CV of the head of the team | D. Overview of external activities which contribute to teaching and/or research |
| VIII. Five core publications | E. Collaborations |
| | F. Valorizable results |

**Table 3. Standard content of the evaluation forms**

| Indicators: | Scale: |
|---|---|
| - Scientific merit of the research / uniqueness of the research, | 9 to 10: High |
| - Research approach / plan / focus / coordination, | 7 to 8: Good |
| | 5 to 6: Average |
| - Innovation, | 3 to 4: Fair |
| - Quality of the research team, | 1 to 2: Low |
| - Probability that the research objectives will be achieved, | Except for "Dominant character of the research", where the experts indicate "fundamental", "applied" or "policy oriented" |
| - Research productivity, | |
| - Potential impact on further research and on the development of applications, | |
| - Potential for transition to or utility for the community, | Ratings reflect the performance of the team as a whole, rather than the academic quality of individual researchers. |
| - Dominant character of the research, | |
| - Reviewer's expertise in the particular research area, | |
| - Overall research evaluation. | |

8. **Special attention for specific research profiles**: Special attention is asked from the experts for innovative and interdisciplinary research. Also the dialogue between experts and research teams plays an important role in a fair treatment of these types of research. These could otherwise be disadvantaged when approached via disciplinary standards or perspectives (Laudel, 2004; Laudel & Origgi, 2006), although such bias is not always observed (Rinia et al., 2001). The productivity in such research areas may for instance suffer from a lower output rate at startup or from a lack of specific journals and needs to be interpreted in its specific context.

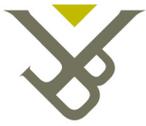

9. **Coherence & Reliability**: In order to obtain coherent results, each expert reviews all teams in the discipline (or as many as he/she feels able to evaluate with reasonable confidence). This creates a uniform reference for the results of the different teams. Coherence is improved by weighting each rating by the reviewer's expertise in the field, as indicated by the expert in each form. The weights for the reviewers' expertise also allow to verify bias with respect to the reviewers' degree of affinity to the research area. Finally, scores on different evaluation aspects can be used as a quantitative indication to verify coherence and reliability (Rons & Spruyt, 2006).

10. **Thorough, yet Timely**: Sufficient time must be dedicated to each phase of the evaluation process in order to ensure the good quality of the evaluation report. Time is needed to find the right experts and to make the draft evaluation files based on centrally stored data. The teams need enough time to complete their files and the experts to review the files. The aim is to finish each evaluation project within one year, so that results and further actions are based on recent data.

11. **Efficient, Well Informed Dialogue**: Evaluation procedures should involve direct dialogue between scientists being evaluated and experts evaluating them (Koetecky, 1999). To avoid misunderstandings, discourage personal predispositions (see 'Independence & Objectivity') and offer more room for nuance, a 'dialogue' format is adopted in the form of a meeting held at the university with the global evaluation coordinator, *all* experts and *all* team leaders, in the presence of the vice-rector for research and the top of the research administration. Discussion sessions are held per research team, each steered by the expert in the particular research area of the team. The global evaluation coordinator moderates the entire program. All team leaders are invited to be present at all sessions, with the exception of the closing session for the experts only. Laboratory visits are organized whenever relevant, depending on the discipline. The evaluation forms sent in by the experts before the meeting are used to produce anonymous overviews of scores and comments per team (ordered by the reviewers' expertise in the area). These overviews provide a starting point for the discussions. They inform the coordinator and the experts of their colleagues' opinions, so that remaining questions can be identified. The experts find the confrontation with team leaders essential for obtaining a clear image. For the teams, it makes a difference to hear the opinions from the experts themselves. The meeting also gives them an opportunity to provide information that is not easily included in the documents, while research administration can situate observations in the current policy context. The absence of such dialogue in a 'remote' peer evaluation would produce several threats to the validity of the results (Gläser & Laudel, 2005).

12. **Concise & Selective Reporting:** Recommendations need to be reported to the appropriate governance levels where remediation can be initiated. This is done via two types of concise reports. A public report describes the context and procedure of the evaluation and contains the global results for the discipline as a whole. It is presented to the Research Council and the Board of the University. Confidential reports per team contain more detailed comments, typically organized in the sections listed in Table 4. While the focus is on the experts' comments, scores indicate the team's relative position with respect to the other teams in their discipline. These reports are given to each team leader by the coordinator during a personal debriefing. The team leaders have the possibility to react to their report. The results per team are also made available to the deans of the concerned faculties and to the rector and vice-rectors. This two-track reporting system allows producing the required public report and at the same time encourages open advice, as details remain confidential (see 'Honest and Unfiltered Opinions').

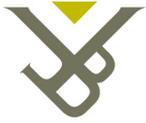

**Table 4. Content of the confidential reports per team**

| Global report: | Confidential reports per team: |
|---|---|
| I. Context (vice-rector for research) | Comments with respect to: |
| II. The discipline and its teams | - Team |
| III. Procedure | - Topics & Innovation |
| IV. Conclusions of the experts | - Output |
| V. Further remarks regarding research evaluation | - Fundamental / Applied / Policy Oriented Research |
| VI. Concluding observations by the coordinator | - Planning |
| ADDENDUM 1: Content of the evaluation files | - Coordination & Collaborations |
| ADDENDUM 2: The evaluation form | - General Conclusions |
| ADDENDUM 3: Coordinator and experts | |
| ADDENDUM 4: Short CV's of the coordinator and the experts | Assessment in numbers, only as an _indication_, showing a team its position within the discipline |
| ADDENDUM 5: Contact information for the teams | |
| | Explanation of the calculations |

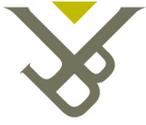

**Practical Execution & Planning**

The central administration working for the Research Council, the R&D department, prepares the planning of the evaluations and organizes and supports all phases of each evaluation project.

Each discipline that is evaluated contains about ten teams. At the VUB, the research teams have a dimension small enough for well-focussed discussions at their level during the meeting. The grouping of teams per discipline widens the scope enough to reveal interesting and hitherto insufficiently exploited or even unexplored links between the groups. As can be seen from footnote 2, some disciplines delimit a particularly wide research area. The actual extent of a discipline in the evaluations is related to its coverage by the research groups active in the domain and to their concentration and degree of specialization. It is observed that the wider scope of certain disciplines often offers unexpected opportunities with respect to innovation, originating at the interface between research areas. More diverse expert backgrounds also lead to different, less expected suggestions for collaboration.

The planning of the evaluations takes into account other events, such as visitations of education programs[4]. Simultaneous evaluations are avoided. Closely following evaluations offer the teams the possibility to optimize the use made of the gathered information. The Research Council can also set priorities based on specific policy interests or according to signals from various data sources. The period of eight years to evaluate all research, imposed by the Flemish government, is in general found to be reasonable. Roughly corresponding to two subsequent PhD periods of 4 years, it is sufficiently large for substantial changes to take place.

In the practical organization of an evaluation project, the following phases can be distinguished:

- **Start up**: Introductory documents describing the evaluation are sent to the teams. Basic data on team composition are verified in order to correctly prepare the evaluation files. Suggestions for experts are collected from the teams. It is verified with the coordinator whether the standard documents require adaptations to the discipline. Changes can for instance be proposed by the research administration based on known specific activity categories for the discipline, in agreement with the evaluated teams.
- **Introductory meeting**: An introductory meeting is organized with the coordinator, the vice-rector for research and the team leaders. Items requiring special attention are already mentioned at this stage.
- **Composition of the expert panel**: For all proposed experts, the coordinator is provided with CV's and information on links found between them and the teams. Standard letters of invitation inform the selected experts.
- **Compilation of evaluation files**: In a first draft, research activities from the preceding five years and background information on personnel and teaching load are collected from central databases. The drafts are sent to the teams for further completion. The efforts of the research administration in making the initial drafts, limits as much as possible the time spent on completion and refinement of the documents by the teams. All evaluation files are given a uniform layout and are sent to the experts together with the evaluation forms.
- **Overview of evaluation forms**: Comments and scores from all returned evaluation forms are organized in overviews per team and in an overview of general comments. These overviews are sent to the experts and the coordinator in order to prepare themselves for the meeting.

- **Evaluation meeting**: An evaluation meeting (typically one entire day) is organized with all team leaders and experts. Notes are taken during all discussions, paying special attention to comments overruling previous ones in the forms.
- **Reports** (global report + detailed reports per team): A draft is made based on all comments (forms and notes) and discussed with the coordinator. The reports are sent to the experts for approval and they are adapted, where necessary, according to their remarks.
- **Debriefing of team leaders**: Personal debriefings are organized so that the coordinator can hand over the results to each team leader and comment them. The teams are given a short period to react, after which the reports become final.
- **Report to Research Council & Board of the University**: The coordinator presents the global report to the Research Council, highlighting the major findings. In the next step, the report is placed on the agenda of the Board of the University and commented by the vice-rector for research.

The main phases and document flows are illustrated schematically in Figure 1. The research evaluations require human resources (personnel, expertise and experience) and sufficiently elaborated specific data processing systems. Besides infrastructure and logistic support the costs of an evaluation project involve expert fees (based on the amount received by EU-referees for research evaluations), travel expenses and hotel accommodation. The investments in infrastructure and human resources do not only serve research evaluation, but also contribute in other ways to research policy (e.g. online public visibility of research activities by mandatory updating of the central database; expertise gained regarding domain specificity that is useful in internal and external research policy).

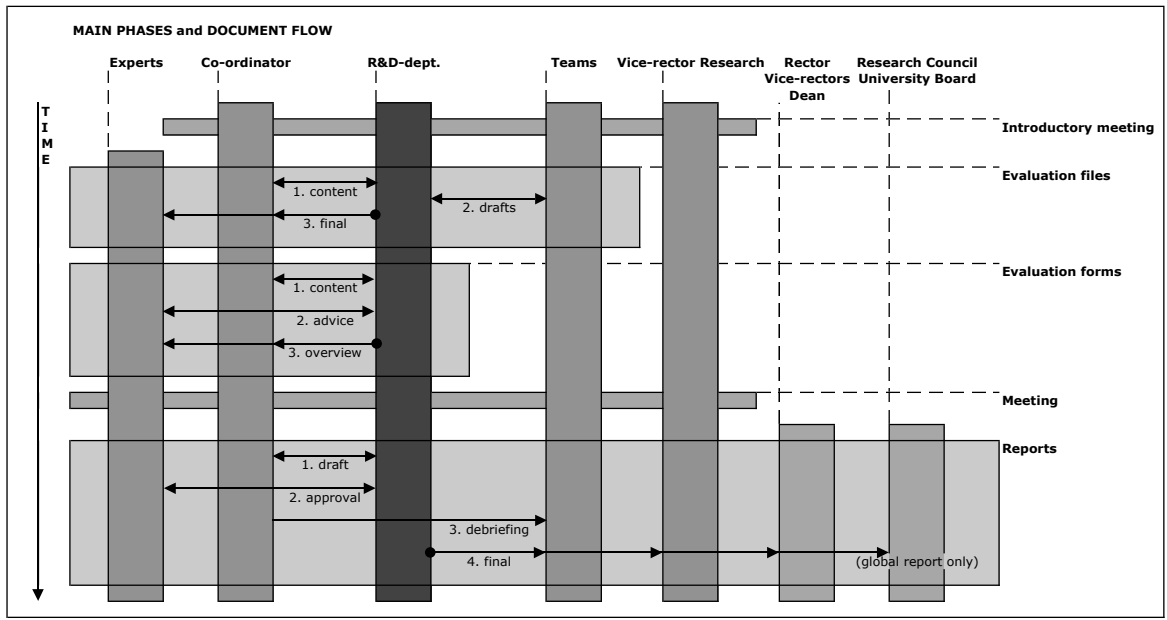

**Figure 1. Main phases and document flow during the evaluation of a research discipline**

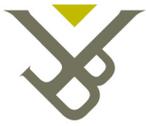

**Current Status, Key Figures & Characteristics**

First two pilot projects (Biotechnology & Law) confirmed that the method could be successfully applied to disciplines in the exact sciences as well as in arts and humanities. In a next step the Research Council approved a standard procedure (Rons, 2001) and planned a complete cycle of evaluations of all disciplines. All research disciplines within the university are evaluated in a cycle of 8 years, the first cycle ending in 2009. Each year two disciplines are evaluated in parallel. Currently half of the disciplines have been evaluated, including disciplines from the exact and applied sciences (BIOTECHNOLOGY, CHEMISTRY[6], ENGINEERING, INFORMATICS, MATHEMATICS, PHARMACY, PHYSICS & GEOSCIENCES) as well as from the social sciences and humanities (ECONOMICS, LAW, PHILOSOPHY & LETTERS, POLITICAL & SOCIAL SCIENCES). Table 5 lists some key figures for the evaluations finalized up to now.

**Table 5. Key figures for the completed research evaluations per discipline**

| | |
|---|---|
| 11 disciplines | 11 expert panels |
| 93 teams | 101 experts from 18 countries |
| ±400 full time equivalent postdoctoral level staff (snap shot) | 550 returned evaluation forms |

Below some characteristics of the quantitative results are shown, including their correlation with results from citation analysis. Since the presentation of quantitative results for the first seven evaluations (Rons & Spruyt, 2006), four more evaluations were finalized and incorporated. A first comparison of the peer review results with quantitative data from the evaluation files is to be presented in 2007 (Rons & De Bruyn, 2007).

Figure 2 shows the distributions of the peer review results per indicator. These distributions are very similar, reaching their maximum at the same value 'good (8/10)', except for one indicator: 'Research approach / plan / focus / coordination'. This indicator corresponds to the nature of most issues repeatedly addressed by the experts in different evaluations (see next section, Table 6).

Significant correlations are found between all pairs of indicators (r=0,59 to 0,93; N=89). Several global aspects of research activities as a whole (e.g. research and team quality, productivity) can indeed be expected to be correlated to some extent. Indicators concerning specific research activities (e.g. publications, projects, conferences) would less likely be correlated.

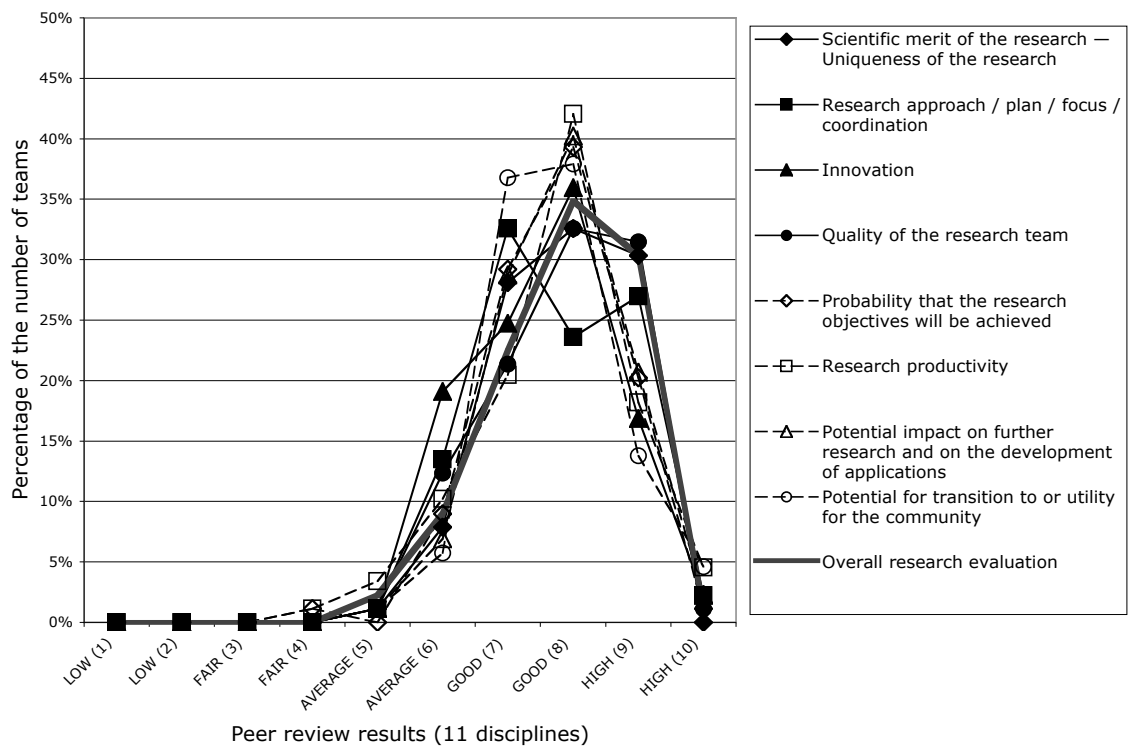

**Figure 2. Relative frequency distribution of peer review results per indicator**

Figure 3 shows the distributions of the peer review results per discipline. These distributions vary in shape, reaching their maxima at values ranging from 'good (7/10)' to 'high (9/10)'. Besides differences in appreciation by the experts of the different disciplines, this variation may also partly reflect differences in reference level depending on the panel.

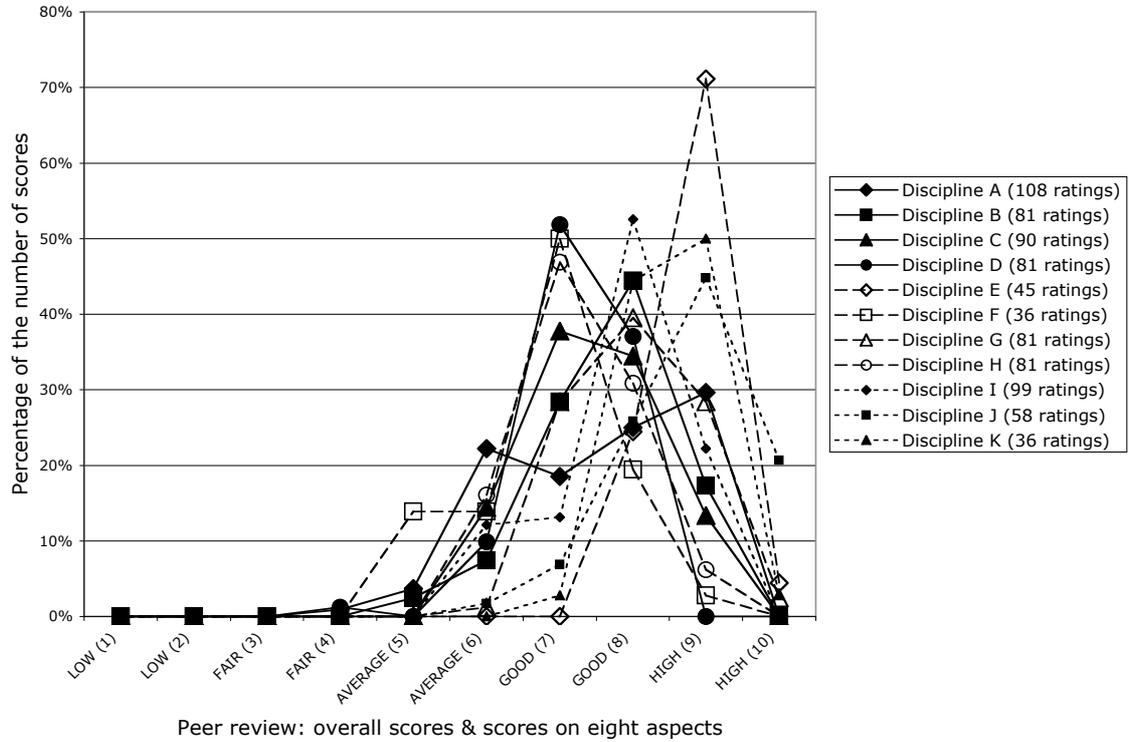

**Figure 3. Relative frequency distribution of peer review results per discipline**

For four disciplines (BIOTECHNOLOGY, CHEMISTRY, ENGINEERING, PHARMACY), the peer review results can be compared to citation analysis results, available for the same teams from a study performed by Visser et al. (2003). This citation analysis includes the well-established ISI-based bibliometric indicators described by Moed et al. (1995). A thorough discussion of these combined results in the framework of reliability and comparability is beyond the scope of this paper. Factors influencing the comparability of results from different methods involve, next to their reliability, also the nature of the indicators, the subject area and characteristics of the methods themselves. Just like peer review, citation analysis has its well-known problems (van Raan, 2005) which need to be taken into account in such comparisons.

Figure 4 shows, as an example, a significant linear correlation between the field-normalized average citation impact (the 'crown indicator' CPP/FCSm) and the peer review score for 'Quality of the Research Team'. Both indicators are size-independent. The figure shows that teams with a high crown indicator value are among the teams that are highly appreciated by peers. The opposite is not always true. Some of the teams that are highly appreciated by peers have crown indicator values below the field average. Reasonable correspondences between results from peer review and citation analysis have been reported for various contexts and aggregation levels and can be expected to some extent. A recent, concise summary and discussion on the comparison of bibliometric results with peer review is given by and Moed (2005). Both kinds of results are never completely independent, because peers will always take certain 'bibliometric aspects' into account. Dependence may even significantly increase when documents handed to the experts include bibliometric reports. Here both results are 'independent' in the sense that the material presented to the peers included mere publication lists for a certain period (including ISI-publications among others), but no citation analysis.

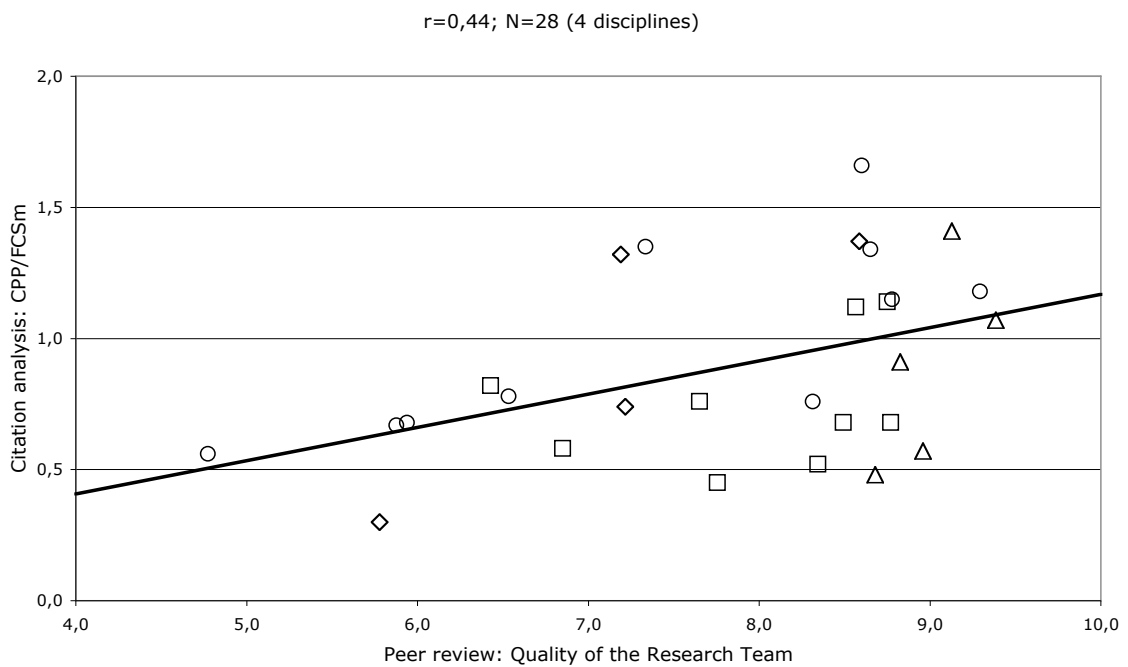

**Figure 4. Correlation of results from peer review and citation analysis**

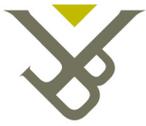

**Qualitative outcomes**

The evaluations generate different types of outcome. Some characteristics of the quantitative results were shown in the preceding section. In this section we focus on the qualitative outcomes. Besides *specific advice* directed to research teams and to the university policy level, there are several other useful outcomes:
- The self-assessment and contacts during the process generate positive effects on the teams' *strategies* and encourage a *group dynamic* within the discipline. This also affects the individual attitudes of researchers in the teams by enhancing the local research culture.
- Many of the experts' remarks implicitly contain more *generally valid recommendations*, which can be collected and made available to the research community as general advice.
- The evaluations offer a better *insight* into the different disciplines and evaluation methods at the university policy level, contributing to the foundation of future policy decisions.

These different forms of outcomes are discussed in more detail in the sections below.

*Specific advice*

The *experts' advice* is handed over confidentially to the research team leaders during a personal debriefing. Each team is expected to take action accordingly, where necessary. This detailed advice is also known to the Deans and the Vice-rector for Research for initiating concerted remediation initiatives.

*Comments directed to the university policy level* may concern specific problems requiring mediation or the recommendation of policy measures. Examples encountered were the physical relocation of teams in a common research environment, a reorganization of teams in larger groups, the relocation of a research team to another department and the provision of technical support in certain experimental research areas involving complex instrumentation. In such cases negotiations can be initiated by the Vice-rector for Research with all parties concerned. Examples of policy recommendations for instance concerned aspects of human resources management (reward systems for a good equilibrium between educational and research efforts, promotion policies based on research output and PhD schools in certain disciplines) and the formulation of expectations and guidelines at institutional level (incentives for international collaboration and publication, reward systems for competitive teams in attracting external funding). These were taken into account in the Research Council's five-year policy plan.

To *take action*, following the evaluation advice, is important to ensure that participation is a useful effort for the teams. It is also a necessary condition to get their full collaboration and to avoid 'assessment fatigue'. Knowing that 'knowledge gained at university policy level really influences decisions' (e.g. regarding input of personnel) is also important to motivate well performing teams to continue their efforts.

*Effects at team level*

Reactions from the teams revealed that the evaluation process generates other positive effects at their level besides the experts' advice and their responses to it:

Writing the paragraphs related to research management implies a *self-assessment*, forcing researchers to look at themselves in a critical way and stimulating them to reflect on research topics, methodology and policy. In the context of the increasing output pressure, some teams would otherwise not give the same priority to such reflection.

The evaluation meeting stimulates communication and interaction between the teams and encourages a *group dynamic*. The teams often get a clearer view on the discipline as a whole and on their own position in it. They get to know better the research activities, capabilities, achievements and problems of the other teams, so that new and enhanced opportunities for collaboration emerge. They also see how other teams manage to cope with similar problems, and learn from each other.

*Generally valid recommendations*

Many of the experts' remarks implicitly contain more generally valid recommendations. The global reports were scanned for more generally applicable positive or negative remarks. These were reformulated as recommendations and organized in a number of categories. The key issues addressed by the experts are listed per category in Tables 6 & 7. The resulting recommendations are made available to the research community as general advice. The fact that such guidelines follow from peers' advice is an important element for their credibility and acceptance by the research community. Of course certain aspects of the reviewers' advice may not be applicable in all situations. The experts themselves are conscious of the fact that their criticisms and recommendations are based on assumptions on how research should be organized and should gain recognition. Certain recommendations stand unquestioned however, regardless of the discipline, such as the requirement of scientific output and its evaluation in a forum of peers. A number of issues addressed by the experts are also addressed in other fora. For example, many issues linked to human resources management are in agreement with CE-recommendations on researchers and their recruitment[5], which the VUB adhered to in an early stage.

**Table 6. Key issues at team level addressed by the experts**

| *Category:* | *Key Issues:* |
| --- | --- |
| team formation & human resources management | efficiency; visibility; continuity; full time staff; perspectives |
| research planning & strategy | strategy, objectives & goals; global vision & research plan; research culture; continuity & sustainability |
| PhD students | postdoctoral / PhD student; recruiting; common structures & activities; 'coaching' & mobility; thesis volume & time; publication |
| publications | journals; peer reviewed; international; balanced output |
| teaching | internationalization; teaching load management |
| internal collaborations | common infrastructure; strategy & policy at discipline level; potential for collaboration |
| international collaborations & networking | opportunities for collaboration; stability; mobility; visibility |

**Table 7. Key issues at institutional level addressed by the experts**

| Category: | Key Issues: |
|---|---|
| human resources management | open competition; expectations & requirements; research time; career stages; sustainability; transition phases |
| research policy | global mission; freedom & guidance; incentives; cost-effectiveness; space & library |

*A tool at institutional level*

The evaluations provide an overview of the different research activities per discipline and a deeper *insight* into the specific nature and problems of each discipline. This knowledge plays an important role as a basis for well-informed policy decisions and the construction of *policy instruments* (e.g. formulation of expectations & requirements per discipline and development of incentives). Within a university research community, the evaluations stimulate a good research culture, which in turn enhances the *competitivity* of the university as a whole.

By organizing thorough research evaluations, the university makes a clear statement on the importance attached to research quality. This contributes to its public *image* and its appreciation in research *management evaluations*. In the evaluation of research management in 2004, the committee approved the VUB's choice for a periodic external research evaluation, thereby demonstrating *"that the researcher and research quality assurance occupy a primary position."* (R.J. van Duinen, 2004, p. 71). The evaluation method also complies with many aspects of *good practice* in internal quality evaluation processes (differentiation between disciplines, communication, transparency, feed back, consequences) as concluded from the Quality Culture Project by the European University Association (*EUA*, 2006).

Qualitative research evaluations can be used to learn more about the *meaning and applicability of quantitative criteria*. A comparison between peer review results and quantitative measures can help understand to what extent the latter relate to research quality. The use of quantitative indicators in international comparisons and national allocation models makes quantitative research output scores a major issue for universities, even when they are aware that these do not necessarily reflect the quality of their research. Experts warn for unwanted side effects on research quality and publication and citation behaviours, when quantitative indicators used by governments at the aggregation level of universities are directly used in intra-university allocation of research funding (Bruneau & Savage, 2002; Butler, 2003; Debackere & Glänzel, 2004). A better understanding of quantitative measures allows the university management level to give the right incentives to their researchers. With this aim in mind, the VUB ordered a bibliometric study (Visser et al., 2003) of the same entities also evaluated in the peer review cycle. This allowed investigating peer review reliability as well as the comparability of results from these different methods (Rons & Spruyt, 2006).

Many uses can be made of the information collected in the framework of the evaluations. Some of them will be addressed in future work.

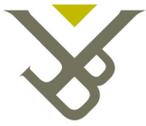

**Conclusions & Future Work**

The method for research evaluation per discipline at the VUB has proven its effectiveness as an instrument to obtain clear and pertinent advice at *research team level* as well as at *university policy level*. Supplementary outcomes, besides this advice, include (i) implicit strategy improving effects at team level, (ii) generally valid recommendations, (iii) better insights into the different cultures, needs and properties of specific research disciplines, and (iv) into different evaluation methods, supporting research policy.

The generated material opens several tracks for future work. The more generally valid recommendations that were derived will, together with other sources, generate gradually refined practical guidelines for research management at different scales and hence at different governance levels. Follow-up studies are planned to learn more about the effects of the evaluations at team level, e.g. (i) an examination of changes in performance following the evaluations (output, funding, collaborations, ...), (ii) an inquiry into the actions undertaken by the teams and (iii) their perception of the most useful elements of the evaluation exercise. The evaluation material also provides a valuable basis for the construction of policy instruments, such as a reference frame per discipline of the specific characteristics of research activities, performance parameters and requirements. In parallel with these derived studies the cycle of research evaluations will be continued.

As a response to the regulations of the Flemish Government imposing a system of internal quality control of the university's research, the choice of a qualitative research evaluation at the VUB has delivered an evaluation method (and the associated tools) that generates a strong multiplication effect through its potential to provide input for many other valuable policy instruments.

**Acknowledgements**

We would like to thank all experts and coordinators who kindly exchanged their views at the occasion of the evaluation projects. Their greatly valued remarks contributed substantially to the present paper and to the validity of recommendations in an international context.

**Notes**

[1] *Decreet betreffende de universiteiten in de Vlaamse Gemeenschap* (Flemish Community, 12 June 1991) and later *Decreet betreffende de herstructurering van het hoger onderwijs in Vlaanderen* (Flemish Community, 4 April 2003).
[2] For the evaluations per research discipline, the Research Council grouped all VUB-teams into the following disciplines: Biology incl. Environmental Sciences; Biotechnology; Chemistry; Economy; Engineering; History; Informatics; Law; Mathematics; Medicine (2 subgroups); Pharmacy; Philosophy & Letters; Physical Education & Kinesitherapy; Physics & Geosciences; Political & Social Sciences; Psychology & Education.
[3] A bibliometric study commisioned by the VUB (Visser et al., 2003) included all disciplines sufficiently covered by Thomson ISI citation indexes for a standard bibliometric study and an extended pilot study for two less well covered disciplines. This involved in total about 60% of VUB-researchers.
[4] By the Flemish Interuniversity Council (VLIR), http://www.vlir.be/, in Dutch.
[5] *COMMISSION RECOMMENDATION on the European Charter for Researchers and on a Code of Conduct for the Recruitment of Researchers* (Commission of the European Communities, CE, Brussels, 11 March 2005).
[6] Evaluation exceptionally involving different experts for each team.

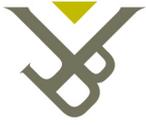